\documentclass[fleqn]{mat01}
\usepackage{times,mathtimy,amssymb,latexsym,,psfig,epsfig}
\begin{document}

\makeatletter
\def\artpath#1{\def\@artpath{#1}}
\makeatother
\artpath{c:/prema/pmcrc}

\setcounter{page}{227}
\firstpage{227}

\renewcommand\thesubsubsection{\!\!\thesubsection.\arabic{subsubsection}}

\providecommand{\norm}[1]{\lvert#1\rvert}

\title{The contact angle in inviscid fluid mechanics}

\markboth{P~N~Shankar and R~Kidambi}{The contact angle in inviscid fluid mechanics}

\author{P~N~SHANKAR and R~KIDAMBI}

\address{Computational and Theoretical Fluid Dynamics Division,
National Aerospace Laboratories, Bangalore 560 017, India\\
\noindent E-mail: pn\_shankar55@rediffmail.com; kidambi@ctfd.cmmacs.ernet.in}

\volume{115}

\mon{May}

\parts{2}

\pubyear{2005}

\Date{MS received 11 August 2004; revised 20 December 2004}

\begin{abstract}
We show that in general, the specification of a contact angle condition
at the contact line in inviscid fluid motions is incompatible with the
classical field equations and boundary conditions generally applicable
to them. The limited conditions under which such a specification is
permissible are derived; however, these include cases where the static
meniscus is not flat. In view of this situation, the status of the many
`solutions' in the literature which prescribe a contact angle in
potential flows comes into question. We suggest that these solutions
which attempt to incorporate a phenomenological, but incompatible,
condition are in some, imprecise sense `weak-type solutions'; they
satisfy or are likely to satisfy, at least in the limit, the governing
equations and boundary conditions everywhere except in the neighbourhood
of the contact line. We discuss the implications of the result for the
analysis of inviscid flows with free surfaces.
\end{abstract}


\maketitle

\section{Introduction}

Consider an inviscid liquid, under a passive, inert gas, partially
filling a smooth walled container. A gravitational field acts on
the liquid. The liquid--gas interface is subject to surface
tension. The interface meets the walls of the container at a line
called the contact line. At any point on the contact line, the
angle between the normal to the gas--liquid interface and the
normal to the solid wall is called the contact angle, $\alpha$. In
the quiescent state, for many pure materials and smooth solid
surfaces, the contact angle for many gas/liquid/solid systems is a
function of the materials alone. Here, we will assume this to be
true and call this contact angle, the static contact angle
$\alpha_{\rm s}$.

The surface tension at the gas--liquid interface, from now on
called the interface, requires that there be a jump in the normal
stress across it if it is not flat. In the inviscid case the jump
is in the pressure and for a static interface the liquid pressure
is just the hydrostatic pressure. Thus the shape of the static
interface depends on a Bond number $Bo$, the ratio of a measure of
the gravitational force to a measure of the force due to surface
tension. But this shape also depends on the static contact angle
$\alpha_{\rm s}$, which provides a boundary condition for the
differential equation that determines the static interface shape.
The static contact angle is therefore a parameter that influences
the static meniscus shape and needs to be prescribed in order to
calculate the meniscus shape.

When the interface is in motion, the surface tension again
requires that a pressure jump, proportional to the interface
curvature and to the coefficient of surface tension, exist across
it. Of course, the pressure in the liquid will now be determined
by the unsteady Bernoulli equation. The situation as regards the
contact angle, however, is more complicated and confused. If one
examines the dynamical equations for the interface it is not at
all obvious that one needs to prescribe a condition on the contact
angle. Moreover, as is well-known, for the case of linearized
disturbances between plane, vertical walls where the initial
interface is plane, the classical solution can be obtained without
any specification of the contact angle, which turns out to be
constant and is equal to $\pi/2$ throughout the motion. On the
other hand, there are many examples in the literature where
analyses and calculations have been made of unsteady potential
motions where a condition on the contact angle has been prescribed
as a boundary condition at the contact line. Just a short list of
these could include Miles \cite{9}, Billingham \cite{3} and
Shankar \cite{15}. It appears that the motivation to prescribe the
contact angle comes from the apparent behaviour of real, viscous
interfaces and the need to tailor inviscid models so that they
lead to realistic results for real interfaces. The question that
we raise here is: are we really free to prescribe the contact
angle in inviscid potential flows and if not, what is the status
of the `solutions' that have appeared in the literature that
purport to model `real' contact angle behaviour?

Our interest in this question is recent and followed an
investigation of contact angle behaviour in viscous flows with
pinned contact lines \cite{17}. We had been aware that Benjamin
and Ursell \cite{2} had shown that the contact angle would remain
constant in linearized, inviscid, potential motions about a flat
interface, i.e. one corresponding to a contact angle of $\pi/2$.
Without carrying out an analysis, we had assumed the result to be
generally true and in \cite{17} had, perhaps influenced by all the
work employing a constant contact angle, even asserted this. When,
however, we recently tried to prove the result, it began to be
clear that there was a problem here: analysis, following \cite{2},
seemed to show that the contact angle {\it cannot}, in general, be
prescribed. Our purpose here is to demonstrate this and to try to
place the existing literature (including ours!), in which a
contact angle condition is employed in potential motions, in
proper perspective. We believe that this is an important issue
because even if an inviscid `solution' is a good model of reality
in some sense, we should be clear in what sense, or approximate
sense, the `solution' is a solution.

\section{Analysis}

We consider the inviscid motion of a liquid in an arbitrary,
three-dimensional smooth walled container. The motion is generated
by the translational motion of the container. The restriction to
translational motions is essential to ensure potential flow in the
moving frame, a necessary condition for some of the results that
will be derived. The fluid is initially in static equilibrium with
the gas above it which is at uniform pressure{\footnote{The
condition of static equilibrium can be relaxed and, in fact, has
to be in the case of time periodic wave motions.}}. The motion is
assumed to start and continue with a uniform pressure over the
interface; we will assume the gas to be passive, i.e. it only
exerts a constant pressure on the liquid interface. In \S2.1, we
write down the equations governing the motion. In \S2.2, we first
consider planar motions; motions in a cylinder of arbitrary
cross-section are examined in \S2.3. Finally, in \S2.4, we
summarize the main results.

\subsection{\it Governing equations}

\begin{figure}\ke
\hskip 4pc {\epsfxsize=10.8cm\epsfbox{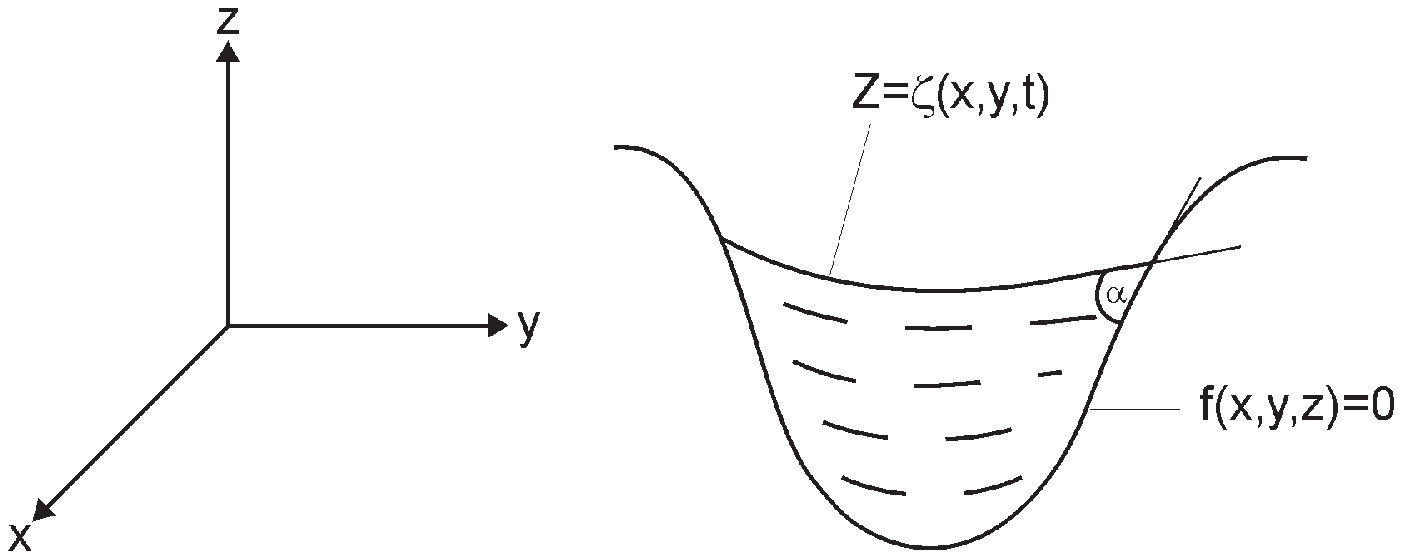}}\vspace{-.5pc}
\caption{A schematic showing a liquid in a container with a free surface
making}
\hangindent=4pc\hangafter=0{\ke contact with the boundary of the container at the contact line.}\vspace{.5pc}
\end{figure}

We write the equations in a reference frame attached to the
container (please, see figure~1). The container wall is given by
$f(x,y,z) = 0.$ Rectangular cartesian coordinates are employed
with gravity generally in the negative $z$-direction. Our analysis
is restricted to the case when the interface is representable by a
single valued smooth function, e.g. by $z = \zeta(x,y,t).$ The
interface motion can be of finite amplitude however. The equations
governing the liquid motion are the continuity and Euler
equations:
\begin{align}
&\nabla \cdot {\bf{u}} = 0,\\[.2pc]
&{\bf{u}}_t + ({\bf{u}} \cdot \nabla) {\bf{u}} = - \nabla p + {\bf{F}},
\end{align}
where $\bf{u}$ is the liquid velocity, $p$ is the pressure and
$\bf{F}$ is the net body force, both real and fictitious. $\bf{F}$
can be an arbitrary function of time. These have to be solved
subject to the boundary conditions on (a) the container wall and
(b) the interface. The condition on the wall is the no-penetration
condition ${\bf{u}} \cdot {{\bf{\nabla }}f} = 0$. The conditions
on the interface $z=\zeta(x,y,t)$ are
\begin{subequations}
\begin{align}
&\zeta_t + u \zeta_x + v \zeta_y = w,\\[.2pc]
&p - p_{\rm a} = - \frac{1}{Bo} \kappa
\end{align}
\end{subequations}
which respectively are the kinematic condition on the interface
and the normal stress condition on it. $Bo$ is a suitably defined
Bond number, $p_{\rm a}$ is the constant ambient pressure over the
interface and $\kappa$ is the local interface curvature. The
interface $z=\zeta(x,y,t)$ is assumed to intersect the container
wall $f(x,y,z)=0$ in a smooth curve called the contact line. It is
further assumed that $\zeta$, ${\bf u}$ and $p$ are analytic in
all variables and that the smoothness of the solutions is upto and
including the boundary. It must be pointed out that the existence
of such solutions is not known at present. The possibly time
dependent angle $\alpha$ made by the contact line with the wall is
given by
\begin{equation}
\frac{f_z - f_x \zeta_x - f_y \zeta_y }{\norm{\nabla f} [1 +
\zeta_x^2 + \zeta_y^2]^{1/2}} = \cos \alpha(t),
\end{equation}
where (4) is to be evaluated at any point
$(x_c(t),y_c(t),\zeta(x_c(t),y_c(t),t))$ on the contact line. We
will show, over the next two sub-sections, that no {\it a
priori} prescription of $\alpha(t)$ is possible though in a few
cases, it turns out to be a constant equal to its initial\break value
$\alpha_{\rm s}.$

\subsection{\it Planar motions}

The container is of arbitrary cross-section in the $x$--$z$ plane
and the motion is 2D. The contact line in this case consists of
just two points (A and B, say). During the course of the motion,
let the contact line traverse regions R of the wall surface. We
will distinguish two cases depending on whether R is locally flat
or not. We designate the former the `flat or straight-wall' case
and the latter as the `curved-wall' case. Examples of the former
are rectangular and wedge-shaped containers (with the $z$-axis
along the wedge); a cylindrical channel is curved-walled. We first
consider the straight-wall case.

\subsubsection{\it The flat or straight-wall case.}

The $x$- and $z$-axes are chosen such that, locally, the body
surface is a line of constant $x$, say $x=0.$ There will be an $x$
component of gravity in this case; the gravity direction has no
bearing on the analysis however. The no-penetration condition
implies $u(0,z,t)=0$ from which it follows that
$u_{z^{(n)}}(0,z,t) = 0$ where $u_{z^{(n)}}$ is the $n$th
derivative of $u$ with respect to $z$.

Now, differentiate (3a) with respect to $x$ to obtain an equation
valid on the interface and hence on the contact line
\begin{subequations}
\begin{align}
&\zeta_{xt}(x,t) + u(x,\zeta(x,t),t) \zeta_{xx}(x,t) +
[u_z(x,\zeta(x,t),t) \zeta_x(x,t)\nonumber \\[.2pc]
&\quad\, + u_x(x,\zeta(x,t),t) ] \zeta_x(x,t) \!=\! w_x(x,\zeta(x,t),t) + w_z(x,\zeta(x,t),t) \zeta_x(x,t).
\end{align}
Using the fact that an inviscid flow starting from rest has to be
irrotational both in an inertial frame and a frame translating with
respect to the inertial, we get $u_z - w_x = 0,$ which with the
continuity equation $u_x + w_z = 0$ allows (5a) to be written in the
form
\begin{equation}
\zeta_{xt} = u_z - (2 u_x + \zeta_x u_z) \zeta_x - u \zeta_{xx}.
\end{equation}
\end{subequations}
where the arguments in (5a) have been dropped to reduce the clutter.
When (5b) is applied at the contact line we obtain
\begin{equation}
\zeta_{xt}(0,t) = - 2 u_x(0,\zeta(0,t),t) \zeta_x(0,t).
\end{equation}
Note that, if the container were to be rotating as well, $u_z -
w_x \ne 0$ and the above analysis would not apply. Equation (6)
shows that $\zeta_{xt}$ at the contact line cannot be specified
arbitrarily; it certainly is not zero in general{\footnote{It
should be noted that we cannot set $u_x(0,\zeta(0,t),t)=0$ and
argue that the contact angle is constant; the reason is that one
cannot specify two conditions at the boundary in a potential flow
and $u(0,z,t)$ has already been set to 0.}}. This means that the
contact angle changes with time in a manner that cannot be
prescribed beforehand. However, for $\alpha = \pi/2,$ not only is
$\zeta_{xt}(0,0)=0$ but also $\zeta_{xt^{(k)}}(0,0) = 0$ for all
$k=2,3, \ldots$. We will show this by mathematical
induction{\footnote{It is not possible to integrate (6) directly
in time to establish the result since all that we know is that
$\zeta_x(0,0)=0$ and hence from (6) that $\zeta_{xt}(0,0)=0$.}}.
Let the induction proposition be
\begin{equation*}
P(k)\hbox{:}\ \zeta_{xt^{(m)}}(0,0) = 0 \,\, \forall \,m =
0,\ldots,k.
\end{equation*}
$P(0)$ is true because $\alpha_{\rm s} = \pi/2.$ Assume $P(k)$ is
true; we will show that $P(k+1)$ is true. Rewriting (5b) as $
\zeta_{xt} = u_z - a(u,\zeta) \zeta_x - u \zeta_{xx},$ the $k$th
time derivative of this equation can be written as
\begin{equation*}
\zeta_{xt^{(k+1)}} = u_{zt^{(k)}} - [a \zeta_x]_{t^{(k)}} - (u
\zeta_{xx})_{t^{(k)}}.
\end{equation*}\newpage
With respect to the above equation, we observe the following:
\begin{enumerate}
\renewcommand\labelenumi{\arabic{enumi}.}
\leftskip -.2pc
\item $\partial \{u_z \} / \partial t^k = 0$ at $x = 0
\forall k = 0,1,2,\ldots$

\item $\partial \{ a \zeta_x \} / \partial t^k = \sum_{m=0}^{k} b_m
a_{t^{(k-m)}} \zeta_{xt^{(m)}} $ where $b_m$ is the binomial
coefficient $C(k,m).$ By the truth of $P(k),$ this sum is zero at
$x=0.$

\item $\partial \{u \zeta_{xx} \} / \partial t^k = \sum_{m=0}^{k}
C(k,m) u_{t^{(k-m)}} \zeta_{xxt^{(m)}}.$ This sum is zero $\forall
\, k = 0,1, \ldots$ as $u$ and all its time derivatives vanish on
the contact line.\vspace{-.5pc}
\end{enumerate}
Thus we have $\zeta_{xt^{(k+1)}}(0,0)=0$ which means $P(k+1)$ is
true. Thus, by mathematical induction, $P(n)$ is true $\forall n =
0,1,\ldots$ and so $\zeta_{xt}(0,t)=0.$ This in turn implies that
$\zeta_x(0,t) = 0$ for all time and the contact angle remains at
$\pi/2$ for all time. Some observations are noteworthy:
\begin{enumerate}
\renewcommand\labelenumi{\arabic{enumi}.}
\leftskip -.2pc
\item Irrotationality of the motion is necessary but not sufficient.

\item This is a nonlinear result, i.e. it holds irrespective of the
perturbation amplitude and the shape of the static meniscus as
long as the initial contact angle $\alpha_{\rm s}$ is $\pi/2.$ In
particular, the static meniscus need not be flat.

\item The result holds as long as the region of the body surface over
which the contact line moves is flat. The shape of the body
elsewhere is immaterial.
\end{enumerate}

\subsubsection{\it The curved wall case.}

While (5b) is still true, the observations following it are not
and nothing can be said about the behaviour of the contact angle
for nonlinear motions even for the case of $\alpha_{\rm s} =
\pi/2.$ It turns out that the contact angle is not constant even
for linearized motions with $\alpha_{\rm s} = \pi/2.$ This is
counter-intuitive as locally one would expect the curved wall to
look `straight' and it is instructive to see where the
straight-wall analysis breaks down in this case. Representing the
body surface by $z = g(x)$ the no-penetration condition is written
as
\begin{subequations}
\begin{equation}
u g' - w = 0.
\end{equation}
Differentiating (7a) on the wall yields
\begin{equation}
g'' u(x,g(x)) + g' \{u_x + u_z g' \} = w_x + g' w_z
\end{equation}
$\alpha_{\rm s} = \pi/2$ yields the relations $\eta'_{\rm s}(x_c)
= - 1/ g'$ and $g''(x_c) = \eta_{\rm s}'' / \eta_{\rm s}^{\prime
2}$, where $\eta_{\rm s}(x)$ is the static meniscus and the primes
denote differentiation with respect to $x$. Using these relations
in (7b) yields
\begin{equation}
\eta_{\rm s}'' u(x_c,z_c) = \eta_{\rm s}^{\prime 2} w_x + \eta_{\rm s}'
(u_x - w_z) - u_z.
\end{equation}
\end{subequations}
The initial interface may be flat or curved depending on the value of
$g'$ (infinite or finite) at the contact line. If (3a) is linearized
about this initial interface we have
\begin{subequations}
\begin{equation}
\eta_t + u \eta'_{\rm s} = w,
\end{equation}
where $\eta(x,t)$ is the perturbation, caused by the container
motion, of the static meniscus $\eta_{\rm s}.$ We need to show
that $\eta_x(x_c,t) = \eta_x(x_c,0)$ for all time. To this end,
differentiate (8a) with respect to $x$ to obtain
\begin{equation}
\eta_{xt} + (u_x + \eta'_{\rm s} u_z) \eta'_{\rm s} + u
\eta''_{\rm s} = w_x
\end{equation}
which on using (7c) becomes
\begin{equation}
\eta_{xt} = (1 - \eta_{\rm s}^{\prime 2}) ( w_x + u_z) + \eta_{\rm
s}' (w_z - 2 u_x).
\end{equation}
\end{subequations}
Thus $\eta_{xt}(x_c,0)=0$ as the fluid is at rest initially.
However, the time derivatives of $u$ and $w$ are in general not
zero which means that $\eta_{xt}(x_c,t) \ne 0$ for arbitrary time
$t.$

\subsection{\it Three-dimensional motions}

An analysis similar to the one in \S2.2 shows that the contact
angle cannot be prescribed arbitrarily in the three-dimensional
case as well. However, following Benjamin and Ursell \cite{2}, we
will now show that a contact angle of $\pi/2$ remains constant for
linear (infinitesimal amplitude) wave motions in a right cylinder
of arbitrary but smooth cross-section standing on one of its ends.
Let the body cross-section be given by $g(x,y) = 0$ and the
interface by $\xi(x,y,z,t) = z - \zeta(x,y,t) = 0,$ then we have the
body and interface unit normals as
\begin{align*}
\hat{n}_{\rm b} &= - \frac{g_x \hat{i} + g_y \hat{j}}{\norm{\nabla g}}, \\[.2pc]
\hat{n}_{\rm i} &= \frac{- \zeta_x \hat{i} - \zeta_y \hat{j} + \hat{k}}{\sqrt{1 + \zeta_x^2 +
\zeta_y^2}},
\end{align*}
whence the contact angle $\alpha$ is given by
\begin{equation*}
\cos \alpha = \hat{n}_{\rm b} \cdot \hat{n}_{\rm i} = \hat{n}_{\rm b}
\cdot \nabla \zeta = \frac{\partial \zeta}{\partial n_{\rm b}}.
\end{equation*}
Since $\alpha_{\rm s} = \pi/2,$ we have initially $\partial \zeta /
\partial n_{\rm b} = 0.$ Linearizing (3a) and making use of the
irrotationality of the motion, the kinematic condition is written as
\begin{equation}
\zeta_t = \frac{\partial \phi}{\partial z},
\end{equation}
where $\phi$ is the velocity potential governing the motion. Note
that (9) is applied on $z=0.$ Differentiating (9) in the direction
of $\hat{n}_{\rm b},$ we have
\begin{equation}
\zeta_{n_{\rm b} t} = \frac{\partial^2 \phi}{\partial n_{\rm b}
\partial z}
\end{equation}
which on using the no-penetration condition on the contact line
$\partial \phi / \partial n_{\rm b} = 0,$ leads to $\zeta_{n_{\rm
b} t} = 0$ on the contact line. Since this is true for all time,
this means $\partial \zeta / \partial n_{\rm b} = 0$ for all time,
i.e., the contact angle remains $\pi/2.$

The same result does not hold in the nonlinear case. We show this
for the simplest case of a right circular cylinder. Employing
cylindrical coordinates $(r,\theta,z),$ the kinematic condition is
\begin{equation}
\zeta_t + u \frac{\partial \zeta}{\partial r} + \frac{v}{r}
\frac{\partial \zeta}{\partial \theta} = w
\end{equation}
applied on $z = \zeta(r,\theta,t)$ and $u,v$ and $w$ are the
$r,\theta$ and $z$ components of velocity. $\alpha_{\rm s} =
\pi/2$ translates to $\partial \zeta /\partial r = 0$ on $r = a$
where $a$ is the radius of the cylinder. Differentiating (11) with
respect to $r,$ we have
\begin{subequations}
\begin{equation}
\zeta_{rt} + u \zeta_{rr} + (u_r + u_z \zeta_r) \zeta_r +
\frac{v}{r} \zeta_{r \theta} - \frac{v}{r^2} \zeta_{\theta} +
\frac{\zeta_\theta}{r} (v_r + v_z \zeta_r) = w_r + w_z \zeta_r.
\end{equation}
Using irrotationality and continuity, the above equation can be
written as
\begin{align}
\zeta_{rt} &= - u \zeta_{rr} - 2 u_r \zeta_r + u_z (1 - \zeta_r^2)
+ \frac{v}{r} \left(\frac{\zeta_\theta}{r} - \zeta_{r
\theta}\right)\nonumber\\[.2pc]
&\quad\,- \frac{v_r \zeta_\theta}{r} - \frac{v_{\theta}
\zeta_r}{r} - \frac{v_z \zeta_\theta \zeta_r}{r}.
\end{align}
Since the motion starts from rest, $u=v=w=0$ initially. So are all
the spatial derivatives of velocity. Initially $\zeta_r$ and
$\zeta_{r \theta}$ are both zero on the contact line. Finally,
$\zeta_{rt}(a,\theta,0) = 0.$ For this to hold for all time, we
need to show that the higher derivatives vanish as well, as we did
in the 2D case. Differentiating (12b) once, we have
\begin{align}
\zeta_{rtt} &= (1 - \zeta_r^2) (u_{zt} + u_{zz} \zeta_t) - 2 u_z
\zeta_r \zeta_{rt} - 2 (u_{rt} + u_{rz} \zeta_t) \zeta_r - 2 u_r
\zeta_{rt} \nonumber\\[.2pc]
&\quad\, - (u_t + u_z \zeta_t) \zeta_{rr} -
u \zeta_{rrt} + \frac{v}{r} \left(\frac{\zeta_{\theta t}}{r} -
\zeta_{r \theta t} \right) + \frac{1}{r}
\left(\frac{\zeta_\theta}{r}-\zeta_{r
\theta} \right)\nonumber \\[.2pc]
&\quad\, \times (v_t + v_z \zeta_t) - \frac{\zeta_{\theta t}}{r}(v_r +
v_z \zeta_r) - \frac{\zeta_\theta}{r}\{ v_{tr} + v_{rz} \zeta_t + v_z
\zeta_{rt}\nonumber\\[.2pc]
&\quad\, + \zeta_r(v_{zt}+v_{zz} \zeta_t) \} - \frac{\zeta_{rt}}{r}
v_{\theta} - \frac{\zeta_r}{r} (v_{\theta t} + v_{\theta z} \zeta_t).
\end{align}
\end{subequations}
It can be shown that $\zeta_{rtt}(a,\theta,0)=0.$ However the time
derivative of (12c) will contain terms like $\zeta_{\theta t}
v_{tr} / r$ which are not necessarily zero and hence
$\zeta_{rt}(a,\theta,t) \ne 0$ for arbitrary time $t.$ It can be
shown however that a contact angle of $\pi/2$ is preserved by the
class of axisymmetric motions (see Appendix~C). This is consistent
with the result obtained in the two-dimensional case.

\subsection{\it A summary of the results}

We summarize, for the convenience of the readers, the main points
of the last two sections. The most important point is that in
inviscid fluid motions starting from rest in a container, the
contact angle cannot be prescribed in advance and neither is there
need for such a prescription. However, if one insists on
prescribing the contact angle, this would necessarily result in a
`weak-type' solution -- one that is in violation of the actual
behaviour of the contact angle. However, in the special case of
$\alpha_{\rm s} = \pi/2,$ there exist situations where the contact
angle remains constant throughout the motion. These situations
include cases of curved static menisci. Prescription of conditions
other than this in this case will again lead to `weak-type'
solutions.

\section{Discussion}

Our discussion will center on the considerable confusion that
exists in the literature for the last five decades, on the role of
the contact angle in inviscid fluid motions. First we should make
clear that there is no confusion whatsoever in the classical
literature on wave motion in liquids, for example as given in Lamb
\cite{8}. In the classical literature, capillarity is considered
only in situations where the liquids are not bounded by solid
boundaries; in such situations there is no contact line and so the
difficulty we are considering does not arise.

\subsection{\it Examples from the literature}

We will substantiate our statements above by presenting below a
small sample of the literature dealing with inviscid contact
lines. We hasten to add that we have no wish to criticize any
particular worker or group; indeed we cite our own work as
manisfestations of this confused state of affairs. The following
is in rough chronological order:

\begin{enumerate}
\renewcommand\labelenumi{(\alph{enumi})}
\leftskip .07pc
\item The article by Reynolds and Satterlee \cite{14} appears in NASA
SP-106, which was the bible for many aerospace engineers for
almost three decades. This excellent article deals with all
aspects of the low-gravity behaviour of liquid propellants. In
dealing with low-gravity sloshing, they indicate the boundary
conditions to be imposed, `plus a contact angle condition'. For
the linearized problem they suggest that this takes the form
`$h_r=\gamma h$' where $h$ is the perturbation to the interface
height and $\gamma$ is a constant to account for `contact angle
hysteresis'. After the formulation of the general linearized
problem, they indicate that they are still faced with a
`formidable problem'; then they just deal with the special flat
interface case in a cylindrical tank. For this special case they
point out ``Note that we were unable to enforce any contact point
condition'' and further note that the contact angle is unchanged
for this solution. It is to be remarked that the authors wished
to impose a contact angle condition but were unable to do so and
found the contact angle to remain constant ($=\pi/2$). The reason
is that they were working with the `classical' spatial
eigenfunctions and could only recover the `classical' solution,
which our analysis showed will lead to a constant contact angle in
the flat interface case.

\item The papers by Moore and Perko \cite{10} and Perko \cite{12} are
important not only because they are among the first to deal with
large-amplitude motions and surface instabilities leading to
breakers, but also because they suggest new methods of dealing
with the liquid sloshing problem. Both papers deal with the
initial value problem for curved interfaces under the influence of
capillarity and are based on expanding the velocity potential in a
series of harmonic functions with time varying coefficients. The
solution method in \cite{10} is such that the evolution of the
interface, at each increment of time, does not involve the current
contact angle. In fact the only role of the contact angle is in
determining the initial interface shape. Consequently, we find
from their interesting figure~3, which shows breakers at the wall,
that the contact angle appears to change with time. These
developments are entirely consistent with the analysis in \S2
which showed that if $\alpha_{\rm s} \neq \pi/2$, the contact
angle will vary in general. Perko \cite{12} extends the earlier
analysis to the general axisymmetric case. However, the author
also points out that it ``includes the constant contact angle
boundary condition {\it necessary to have a well-posed problem in
computation}'' (emphasis added). This is puzzling since
computations were possible in the earlier paper without any
condition on the contact angle, and in fact it was not possible to
impose such a condition. The author points out that the new
figure~4, with a constant contact angle, is for the same
conditions as for the earlier figure~8, where it was not constant;
there is no suggestion as to which one we should prefer or why. It
appears from \S2 that the `solutions' in Perko \cite{12} are not
classical ones.

\item Chu \cite{4} is representative of a large class of papers in the
1960s and 1970s which suggested ways in which slosh frequencies,
forces and moments could be calculated for axisymmetric containers
under low-gravity conditions. This particular paper suggests the
use of a Galerkin type of procedure. As regards the contact line,
Chu says ``In addition, there is an interface contact point
condition which takes the form \dots'' and gives the condition
given in (a) above.

\item It was pointed out in \S2 that if the initial contact
angle is $\pi/2$ and the side walls are straight, even a
non-linear, two-dimensional motion would maintain the contact
angle at $\pi/2$. This implies that if non-linear,
capillary-gravity standing waves exist, every such wave would be a
solution of a problem where straight walls are located at the
nodal points of the wave motion. An example of this possibility is
the solution found in Concus \cite{5} where such a solution is
found to be third order in the amplitude of the waves.

\item Myshkis {\it et~al} \cite{11} is an encyclopaedic book on
low-gravity fluid mechanics. However, in Chapter~5 when they formulate
the small oscillation problem, after writing down eq.~(5.2.14) all
they have to say on the contact angle is ``\dots(5.2.14) is the
linearized condition for the conservation of the contact angle''.
In fact, the classical theory tells us it will not be conserved in
general.

\item A totally different aspect is presented by Benjamin and Scott
\cite{1}, who appear to have been the first to consider the
natural frequencies of a confined liquid with a flat interface,
but whose contact line is pinned. The inviscid modelling of this
situation is immediately seen to be problematic because the
no-slip condition seems to be required at the contact line, a
condition that cannot in general be satisfied by a potential flow.
In other words, there is no classical solution to this problem. In
light of this, Benjamin and Scott using the framework of
functional analysis, define various function spaces, operators and
other tools to formulate a `weak solution' to the key equation
(7b) of their paper and finally get estimates for the frequencies
using Rayleigh's principle. Moreover, they show that their
theoretical estimates agree well with measured values of the wave
periods.

\item Hocking \cite{7} is a widely cited paper because a new model for
the contact angle condition, `${\partial \eta'}/{\partial t'} =
\lambda' {\partial \eta'}/{\partial n}$', is introduced for the
$\alpha_{\rm s}=\pi/2$ case; here $\eta'$ is the surface elevation
of the small disturbance above the flat static interface. The
model is introduced to account in some way for the `wetting
property' of the fluid and includes both the free and pinned edge
conditions as limiting cases. In his introduction Hocking says
``The presence of capillarity adds an extra term to the
free-surface pressure condition. ........ The increase in the
order of the pressure condition, however, {\it requires extra
conditions to be imposed when the solution is sought in a finite
region}'' (emphasis added). This is a common misconception. While
it is true that capillarity increases the order of the equation
governing the {\it static meniscus} and hence the number of
boundary conditions needed in this case, this does not apply in
the dynamical situation. This can be seen by just considering the
flat interface case between vertical walls. In any case, the
analysis of \S2 shows that no extra contact angle condition can be
prescribed for classical\break solutions.

\item It suffices now to mention that the old confusions persist
into the new millenium, typical samples being \cite{3,15,16,19}.
\end{enumerate}

\subsection{\it The current status of the contact angle in
inviscid flows and how it has come about}

We will now summarize how the dynamical, inviscid contact angle
appears to be viewed by most workers and why it has come to be
this way. Recall that in the classical works the question of the
contact angle never arose because confined flows with boundaries
and capillarity were not normally studied. Early studies on the
latter were confined to linearized, two-dimensional flows between
vertical walls; here the classical, exact solutions correspond to
a contact angle of $\pi/2$, which is maintained throughout the
motion. In the 1960s the space programmes required solutions to
more general problems involving curved static interfaces and
static contact angles other than $\pi/2$. The difficulty posed by
these problems forced approaches that were either semi-analytical
or numerical and some like Moore and Perko \cite{10} did not
attempt to impose a dynamical contact angle condition;
$\alpha_{\rm s}$ affected the initial conditions alone through the
static meniscus. The imperative to impose a contact angle
condition at the contact line, not permitted in general by the
classical inviscid formulation, appears to have come from
experimental observations of real, viscous contact lines. It is
well-known \cite{6,16,13,18} that real, dynamic viscous contact lines
display complex behaviour and are not at all well-understood with
many parameters playing a role. It is in attempting to model this
complicated behaviour in an inviscid framework that the need for
contact angle conditions began to be felt and then applied. The
earlier models of a constant contact angle and the one used in
\cite{14} (essentially to model contact angle hysteresis) are in a
sense non-dynamic. Hocking's \cite{7} model is a dynamic one,
attempting in an inviscid framework to account for contact line
hysteresis or viscous wetting effects at the contact line. In any
case, the purpose is to account for viscous and other real effects
in an inviscid, potential model of the flow.

Thus the need to more realistically model the dynamic contact line
appears to require the freedom to {\it impose} a condition at the
inviscid contact line. But as was shown in \S2 the classical
field equations and boundary conditions do not in general provide
this freedom. Then the natural question is, what is the status of
the very large and important body of work in which a contact angle
condition is imposed, in violation of the classical formulation?
Let us call solutions which are obtained without such a condition
`classical'. Then this body of work referred to does not deal
with classical solutions. This means that these `solutions' will
be found to violate at least some of the boundary conditions or
assumptions at the contact line.

\subsection{\it The importance of the present result for inviscid
free-surface flows}

A natural question that arises is: how important is the present
result that the contact angle cannot, in general, be prescribed in
a potential flow? Before we attempt to answer this question we
would like to consider the situations shown in figure~2.

The configurations in (a) and (b) are planar while (c) is
axisymmetric and $\alpha_{\rm s}=\pi/2$ in all three cases.
According to our theory, finite amplitude motions in (a) preserve
the contact angle, while the contact angle is not preserved in (b)
and (c) even for infinitesimal motions. This is not a result that
is obvious at all and shows that we must be very careful when
dealing with the contact angle in inviscid flows.

\begin{figure}\ke
\hskip 4pc {\epsfxsize=7cm\epsfbox{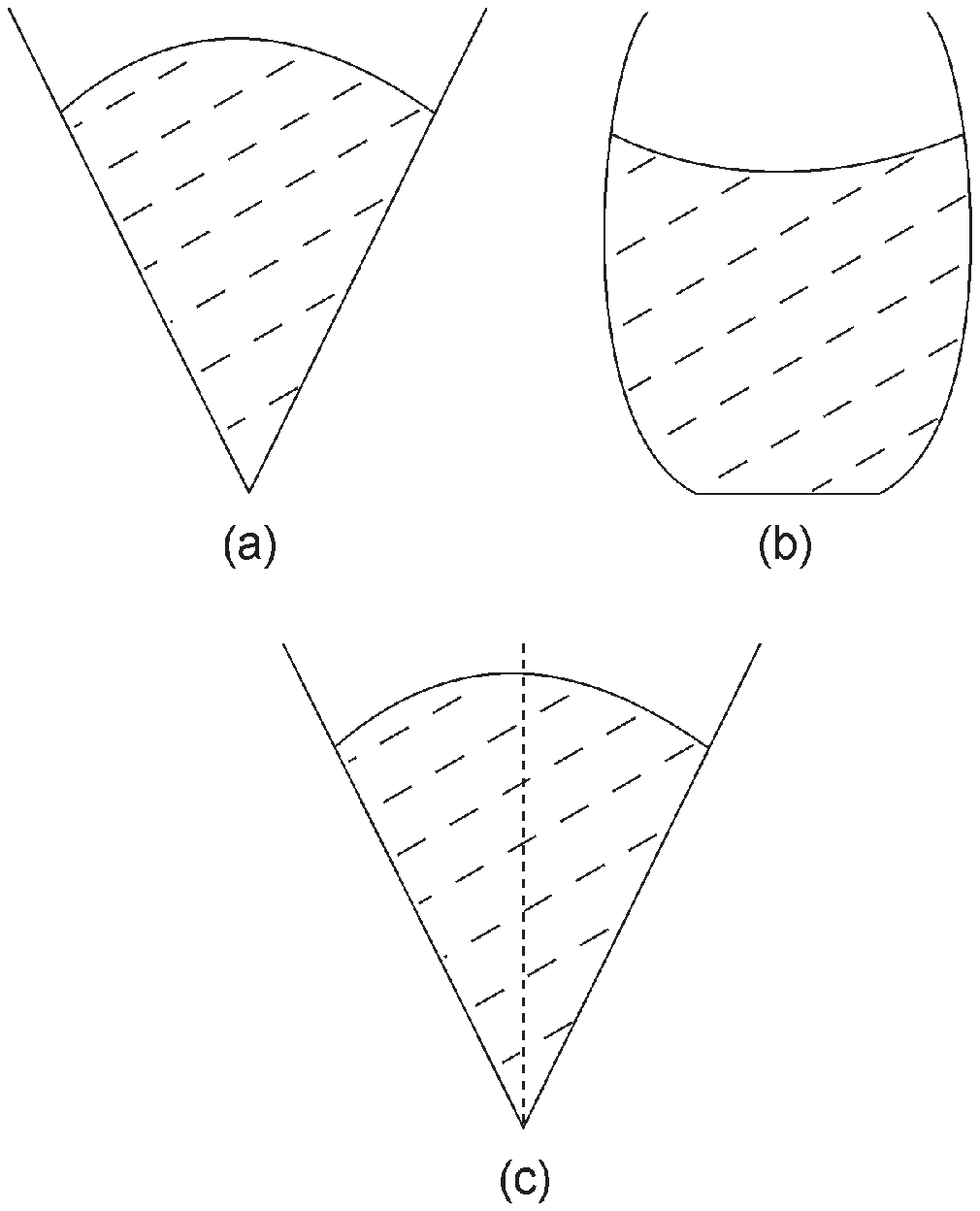}}\vspace{-.5pc}
\caption{The \,figure \,shows \,three \,examples \,\,of \,liquids \,in \,\,containers \,which
\,have}
\hangindent=4pc\hangafter=0{\ke gas--liquid interfaces above them. In
all the three cases the static contact angle $\alpha_{\rm s} =
\pi/2$. The geometry is planar in {\bf (a)} and {\bf (b)}, while in {\bf
(c)} the container is a cone with the same half angle as in {\bf (a)},
which too is flat walled. The present theory predicts that the contact
angle will be preserved in {\it finite amplitude motions} in {\bf (a)} but
will not be preserved even in {\it infinitesimal motions} in {\bf (b)} and
{\bf (c)}.}\vspace{.4pc}
\end{figure}

Both from general considerations and from the above example it
should be clear that the result is of real importance in basic
inviscid flow theory. But one may still inquire whether the result
is of any practical importance, for example, in the calculation of
the natural frequencies of surface waves or in sloshing
calculations. This is not easy to answer because we do not know
how to calculate such things {\it without making any assumptions
about the contact angle}. For example, say we wish to find the
natural frequency in a case where the static contact angle $\alpha
\ne \pi/2$. It appears that the frequency cannot come out of an
eigenvalue problem by assuming a solution harmonic in time and an
expansion in terms of spatial modes $\phi({\bf x})$ because the
latter would imply some assumption about the contact angle at the
boundary. It appears that the only way out would be to do the
initial value problem with the correct static meniscus and let the
field evolve freely. But it is unlikely that this can be done
easily and certainly not analytically. However, we should point
out that Moore and Perko \cite{10} solve the initial value problem
without making any assumption on the contact angle and apparently
without making any assumption on a functional form for the
interface. On the other hand, Perko \cite{12} solves the same
problem holding the contact angle fixed at the static value. His
figure~4 can be compared directly with the earlier figure~8 of
Moore and Perko. While there are differences, fortunately and as
might be expected, the qualitative overall pictures are similar.
This suggests that the violation of the correct condition at the
contact line will lead to violation of the interface conditions at
the contact line and except in extraordinary circumstances will
not have much of an effect on the overall field. Thus it appears
that natural frequencies and sloshing modes will not be greatly
affected provided the containers are sufficiently large. This
must be especially true of the natural frequencies which are
related to integrals over the whole field. Indirect evidence in
support of this position is that most frequency calculations done
by different methods, which presumably violate the correct
conditions differently, agree well with one another.

\section{Conclusion}

We have shown that the classical field equations governing the
motion of a confined inviscid liquid under a passive gas do not,
in general, permit the independent specification of a contact
angle condition at the contact line. In fact the only cases where
such a condition may be permissible are when the static contact
angle is $\pi/2$, the container walls are flat, at least in the
neighbourhood of the contact line, and (i) the motion is
two-dimensional or (ii) the motion is a small three-dimensional
disturbance from a flat initial interface. The restrictions are
indeed surprising as is the difference between two- and
three-dimensional motions.

These results have a somewhat serious bearing on the vast
literature that exists in which `solutions' have been found to
inviscid motions in which various contact angle conditions have
been imposed. It is our contention that these cannot be classical
solutions to the classical field equations since classical
solutions do not permit the imposition of a contact angle
condition. It is suggested that these `solutions' belong to an
improperly defined class of `weak-type solutions', in the sense
that they attempt to solve the field equations in an approximate
sense, with some of the equations being solved exactly. The need
for such `solutions' is driven by the compulsion to try to model in
an inviscid framework, the complicated behaviour of moving
viscous contact lines. Examples were given of other cases where a
similar situation exists.

Finally, we have shown that while the present result is of basic
importance in the theory of inviscid free-surface flows, it is
unlikely to seriously affect the practical calculation of natural
frequencies and sloshing modes in containers.

\section*{Appendix A}

In \S2, the initial contact angle $\alpha_{\rm s} = \pi/2$ was
shown to remain constant under certain conditions. One of these
conditions was that the initial interface be the static meniscus
itself and that the fluid be quiescent initially, i.e., we start
from a static equilibrium. This would be the normal procedure of
posing an initial value problem. On the other hand, we can seek
special solutions such as periodic motions where though initially
the fluid is at rest and the pressure over the interface constant,
the state is not one of static equilibrium and the interface is
not the static meniscus. Examples are the well-known linear and
nonlinear periodic oscillations \cite{8,5} between two parallel
vertical walls. The initial contact angle $\alpha_{\rm s} = \pi/2$
would remain constant in this case as well.

\section*{Appendix B}

We consider here the case of $\alpha_{\rm s} = 0$; at first
glance, it seemed the contact angle would be preserved in this
case as well. Since $\eta_{\rm s}'(0,t) = \infty,$ we reorient the
coordinates such that $x$ is along the wall, $z$ normal to it and
$z = \zeta(x,t) = \eta_{\rm s}(x) + \eta(x,t)$ gives the
interface, as before. The initial interface, close to $z=0$ can
also be described as $x = \xi_{\rm s}(z),$ a description that will
be needed below.

\setcounter{equation}{0}
\renewcommand\theequation{B\arabic{equation}}
The linearized kinematic condition is written as
\begin{equation}
\eta_{tx} = (1 - \eta_{\rm s}^{\prime 2}) u_z - \eta_{\rm s}'' u -
\eta_{\rm s}' u_x,
\end{equation}
where $\eta_{\rm s}'(0)=0.$ $\eta_{tx}(0,0)$ is zero as the
initial state is one of rest. The $k$th derivative of (B1) is
\begin{equation}
\eta_{xt^{(k+1)}} = (1 - \eta_{\rm s}^{\prime 2}) u_{zt^{(k)}} -
\eta_{\rm s}'' u_{t^{(k)}} - \eta_{\rm s}' u_{xt^{(k)}}.
\end{equation}
It is easy to see that the first and third terms vanish at $x=0$
as $u_z = \eta_{\rm s}' = 0$ there. If we can show that the second
term vanishes as well, we would be done. In the second term,
$u(0,0) \ne 0$ so we want $\eta_{\rm s}''(0) = 0.$

It is convenient to write the equation for the static meniscus
with gravity along the $x$-axis. The equation would be
\begin{equation}
\xi_{\rm s} - \frac{1}{Bo} \frac{\xi_{\rm s}''}{(1+\xi_{\rm
s}^{\prime 2})^{3/2}} = \lambda,
\end{equation}
where the prime denotes differentiation with respect to $z$ and
$\lambda$ is a non-zero constant equal to $- \kappa(0)/Bo$ where
$\kappa(0)$ is the curvature of the interface at the contact line
and is given by
\begin{equation}
\kappa(0) = \frac{\xi_{\rm s}''(0)}{(1+{\xi_{\rm s}(0)}^{\prime
2})^{3/2}}.
\end{equation}
Now, we can write, by chain differentiation,
\begin{equation}
\eta_{\rm s}''(x) = - \frac{\xi_{\rm s}''(z)}{\xi_{\rm s}^{\prime
3}(z)}.
\end{equation}
Since $\xi_{\rm s}'(0)$ is infinite but $\kappa(0)$ finite and
non-zero, we have from (B4) that $\xi_{\rm s}''(0)/\xi_{\rm
s}^{\prime 3}(0)$ is a non-zero finite quantity. But by (B5), this
means $\eta_{\rm s}''(0)$ is non-zero as well. Hence,
$\eta_{xt^{(k+1)}} \ne 0$ in general and the contact angle will
vary during the motion.

\section*{Appendix C}

In \S2, it was shown that the contact angle will, in general,
change for nonlinear motions in a right circular cylinder.
However, we prove here that the contact angle $\alpha_{\rm s} =
\pi/2$ is preserved for the special class of axisymmetric
nonlinear motions in a right circular cylinder. The proof by
induction follows the general pattern of the proof in \S2.2.1.
Noting that axisymmetric motions mean that the interface is given
by $ z = \zeta(r,t), $ let the induction proposition be
\begin{equation*}
P(k)\hbox{:}\ \zeta_{rt^{(m)}}(0,0) = 0 \,\, \forall \,m = 0,\ldots,k.
\end{equation*}
$P(0)$ is true because $\alpha_{\rm s} = \pi/2.$ Assume $P(k)$ is
true; we will show that $P(k+1)$ is true. Writing (12b) for this
case, we have
\setcounter{equation}{0}
\renewcommand\theequation{C\arabic{equation}}
\begin{equation}
\zeta_{rt} = - u \zeta_{rr} - 2 u_r \zeta_r + u_z (1 - \zeta_r^2)
\end{equation}
as the rest of the terms are identically zero due to axisymmetry.
The $k$th derivative of (C1) will have the following terms:
\begin{enumerate}
\renewcommand\labelenumi{\arabic{enumi}.}
\leftskip -.2pc
\item $\partial \{ u \zeta_{rr} \} / \partial t^k = \sum_{m=0}^{k}
C(k,m) u_{t^{(k-m)}} \zeta_{rrt^{(m)}}. $ This sum is zero
$\forall \, k = 0,1, \ldots$ as $u$ and all its time derivatives
vanish on the contact line.

\item $\partial \{ u_r \zeta_r \} / \partial t^k = \sum_{m=0}^{k}
C(k,m) u_{rt^{(k-m)}} \zeta_{rt^{(m)}}. $ By the truth of $P(k),$
this sum is zero at $r=a.$

\item $\partial \{ u_z (1 - \zeta^2_r )\} / \partial t^k =
\sum_{m=0}^{k} C(k,m) u_{zt^{(k-m)}} (1 - \zeta^2_r)_{t^{(m)}}. $
This sum is zero $\forall k = 0,1, \ldots$ as $u_z$ and all its
time derivatives vanish on the contact line.
\end{enumerate}
Thus we have $\zeta_{rt^{(k+1)}}(0,0)=0$ which means $P(k+1)$ is
true. Thus, by mathematical induction, $P(n)$ is true $\forall n =
0,1,\ldots$ and so $\zeta_{rt}(0,t)=0.$ This in turn implies that
$\zeta_r(0,t) = 0$ for all time and the contact angle remains at
$\pi/2$ for all time.

\end{document}